\documentclass[prb, twocolumn, amsmath, showpacs, amssymb, superscriptaddress, preprintnumbers]{revtex4}
\usepackage{graphicx}
\usepackage{subfigure}
\usepackage{dcolumn}
\newcommand{\wcm}{{ W-cm$^{-2}$}}
\begin{document}
\title{Extended excitons and compact helium-like biexcitons in type-II quantum dots}
\author{{Bhavtosh Bansal}}\email{bhavtosh.bansal@gmail.com}
\affiliation{INPAC-Institute for Nanoscale Physics and Chemistry,
Pulsed Fields Group, KU Leuven, Celestijnenlaan 200D, Leuven
B-3001, Belgium}
\author{S. Godefroo}
\affiliation{INPAC-Institute for Nanoscale Physics and Chemistry,
Pulsed Fields Group, KU Leuven, Celestijnenlaan 200D, Leuven
B-3001, Belgium}
\author{M. Hayne}\email{m.hayne@lancaster.ac.uk}
 \affiliation{Department of Physics, Lancaster
University, Lancaster LA1 4YB, UK}
\author{G. Medeiros-Ribeiro}\email{gmedeiros@lnls.br}
\affiliation{Laborat\'orio Nacional de Luz S\'incrotron, P.O. Box
6192, 13084-971 Campinas-SP, Brazil}
\author{V. V. Moshchalkov}\email{Victor.Moshchalkov@fys.kuleuven.be}
\affiliation{INPAC-Institute for Nanoscale Physics and Chemistry,
Pulsed Fields Group, KU Leuven, Celestijnenlaan 200D, Leuven
B-3001, Belgium}
\date{\today}
\begin{abstract}
We have used magneto-photoluminescence measurements to establish that InP/GaAs quantum dots have a type-II (staggered) band alignment. The average  excitonic Bohr radius and the binding energy are estimated to be 15nm and 1.5 meV respectively. When compared to bulk InP, the excitonic binding is weaker due to the repulsive (type-II) potential at the
hetero-interface. The measurements are extended to over almost six orders of
magnitude of laser excitation powers and to magnetic fields of up to 50 tesla. It is shown that the excitation power can be
used to tune the average hole occupancy of the quantum dots, and hence the strength of the electron-hole binding. The diamagnetic shift coefficient is observed to drastically reduce as the quantum dot ensemble makes a gradual transition from a regime where the emission is from (hydrogen-like) two-particle excitonic states to a regime where the emission from (helium-like) four-particle biexcitonic states also become significant.
\end{abstract}
\maketitle
\section{Introduction}
Until recently---but for a few exceptions---the study of quantum dot (QD)
heterostructures with the staggered (type-II) band alignment had
been largely ignored because of the absence of confinement of one
of the two types of carriers and their presumed poor radiative efficiency. However,
it has come to be recognized that these structures are interesting, especially for their rich physics of excitons.\cite{ribeiro,bansal phase transition, Nakaema, manus apl, biexciton apl, Madureira, degani}

In contrast to the  usual type-I QDs (e.g. InAs/GaAs) where the confinement energy
scale is far greater than the energy of the Coulomb interaction, the role of confinement
in type-II QDs is largely limited to defining the
geometry of the system. This in itself has interesting consequences. The multiply-connected topology can give rise to an oscillatory ground state energy for the magneto-excitons.\cite{ribeiro,  degani} Secondly, type-II QDs also act as nanocapacitors\cite{Geller APL} which selectively accommodate only one type of particles; but once charged they can bind the
complementary particle to form an exciton. The strength of the Coulomb interaction can be modified by screening or magnetic
field.\cite{bansal phase transition} At higher excitation
powers, they can be doubly-charged and form
four-particle bound states (biexcitons). The biexcitons in a
type-II QD system are very unlike their
counterparts\cite{cade} in type-I QDs and quantum
wells. They always have negative `binding energy'.\cite{biexciton apl}
In the atomic physics language, while the usual biexciton is
structurally analogous to a hydrogen molecule, the
biexciton\cite{biexciton apl} in type-II QDs is more like a helium
atom.\cite{manus apl}

In this article we have probed the nature of the ensemble photoluminescence (PL) emission from a sample with InP QDs in a GaAs
matrix. Although the band offsets\cite{pryor-pistol} of {\em bulk} InP and GaAs and some previous studies suggest that this material combination forms type-II structures with electrons localized within the InP quantum dots and free holes in the GaAs matrix, the
energy gap of InP and GaAs is within $100$meV of each other, and the conduction band offset is relatively small.\cite{pryor-pistol} Thus alloying and anisotropic strain within the QDs,\cite{sample growth} can modify the energy gap and the relative offsets in way that is dependent on
the details of the size and shape of the QDs. For example, a comprehensive k.p calculation \cite{pryor-pistol} does not find a type-II alignment in this system. Secondly, thick ($>100$ nm) InP heteroepitaxial layers on GaAs and even homoepitaxial InP have shown a broad
emission peak at $1.34$ eV\cite{InP DAP example} due to donor-acceptor-pair (DA) recombination, which is rather close to the PL energy of InP/GaAs QDs. Furthermore, the  DA emission has characteristics\cite{InP DAP example-blue shift} generally used to classify spatially-indirect excitons from type-II QDs---excitation power-dependent
blue-shift,\cite{wang-apl, Nakaema} spectrally broad emission at a sub-bandgap energy in macro-PL, and narrow emission lines from localized
states in micro-PL.\cite{Nakaema} One of the aims of this study is to unambiguously establish the type-II band alignment in this system and highlight the role of heterostructure boundary conditions on the size and the binding energy of excitons.
In this work we will not discuss the Aharonov-Bohm-type effects associated with the topology of the wave function as these have already been extensively discussed in literature\cite{ribeiro, degani} and manifest on energy scales an order of magnitude smaller than we are interested in here.

Secondly, by extending the measurements to over six orders of magnitude of excitation powers, we have been able to change the average electronic occupancy within the QDs. This leads to significant changes in the diamagnetic shift coefficient that are a result of the interesting physics of biexcitons.
\section{Experimental}
The sample was grown by metal-organic vapor phase epitaxy on (001) GaAs substrate.\cite{sample growth} The QD density was
about $3\times 10^{10} \textrm{cm}^{-2}$ with an average diameter of $32$ nm ($\pm 6$ nm) and average height of about $4$ nm ($\pm
2$ nm) as measured by cross-sectional transmission electron microscopy.\cite{sample growth} But based on experience from other heterostructures, it is possible that the actual size of quantum dots is much smaller.\cite{timm}

Non-resonant (excitation wavelength=532nm) PL measurements were performed at liquid helium temperature with the excitation powers varied between ($\sim 10^{-4}-100$Wcm$^{-2}$). The light from the excitation laser and the sample PL was fibre-optically coupled in and out of (i) a variable-temperature cryostat within the bore of a superconducting magnet (B$\leq 12T$) for magneto-PL measurements at low excitation powers and (ii) a liquid helium bath cryostat whose tail was within the 18mm bore of the pulsed field coil ($B\leq 50T$) for measurements at higher powers $P>30 Wcm^{-2}$. The PL spectra were recorded by an electron multiplying charged-coupled device after being dispersed by a 30cm imaging spectrograph.
Pulsed magnetic fields of up to 50 T were generated using a 5 kV, 500 kJ capacitor bank. The field had a duration of about 20ms, during which several PL spectra were recorded. Fig. 1(inset) shows three representative spectra under different conditions.

\begin{figure}[!t]
\begin{center}
\includegraphics[width=9.3cm,
  keepaspectratio,
  angle=0,
  origin=lB]{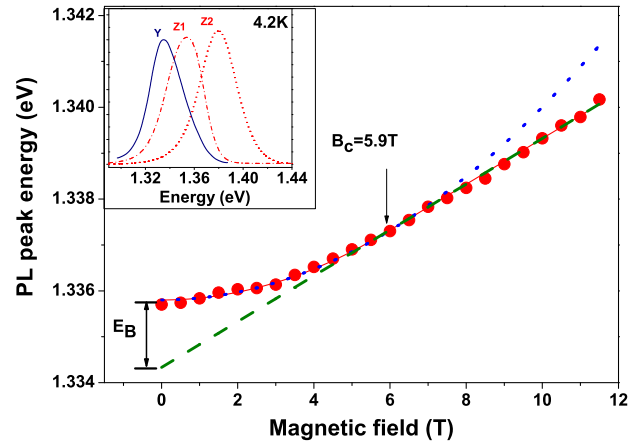}
\caption{\label{fig: Fig1}{(Color online) (Inset) Three representative spectra from the InP/GaAs QD ensemble taken under different conditions. Y:magnetic field B=0,  laser power P$\sim 10^{-4}$\wcm, Z1:B=0, P$\sim 85$\wcm, Z2:B=50T, P$\sim 85$\wcm. The relative amplitudes have been arbitrarily scaled. (main figure) The measured peak-shift as a
function of magnetic field at the excitation power of about $10^{-4}$\wcm. (solid line) Fit to Eq.\ref{eqn:
FIPS}. (dotted lines) Eq.\ref{eqn:
FIPS} (a) and \ref{eqn:
FIPS}(b) plotted separately. The
crossover from Coulombic confinement to magnetic confinement
occurs at 5.9 tesla and corresponds to an effective Bohr radius
of $15$ nm. The extrapolation of the linear slope measured at
high field to zero field gives the binding energy, $E_B=1.5$ meV.}}
\end{center}
\end{figure}
\section{Results and Discussion}
\subsection{Excitons: Type-II band alignment}
Fig.\ref{fig: Fig1} shows the magnetic field-dependent PL peak-shift measured at
$4.2$K, at a very low excitation power of about $10^{-4} Wcm^{-2}$. Using the model of a hydrogenic exciton, one can semi-phenomenologically describe this shift by the following equations:\cite{manus apl, bansal phase transition}
\begin{eqnarray}\label{eqn: FIPS}
E(B)=E_0+ {q^2\langle\rho_x^2\rangle\over 8 \mu}B^2,\,\,\,
\textrm{for} B<B_c \nonumber
\;\;\;\;\;\;\;\;\;\;\;\;\;\;\;(1\textrm{a})\\
E(B)=E_0- {\hbar^2\over 2\mu\langle\rho_x^2\rangle}+
{\hbar q\over 2\mu }B, \,\,\, \textrm{for} B>B_c \nonumber
\,\,\,\,\,\;\;\;(1\textrm{b})
\end{eqnarray}
These equations are derived under the assumption that the magnetic field-induced change in the ground state energy of a hydrogenic exciton from the low-field regime of quadratic (diamagnetic) shift to the high field linear shift (due to transitions between effectively free Landau levels) is adiabatically continuous (i.e. its functional form is continuous and differentiable) between two well-defined limits. Note that the high field limitis approximate as it assumes that the transitions are between {\em free} Landau levels and ignores the weak ($\log\,^2B$, as $B\rightarrow \infty$) dependence\cite{Landau_Level_Spectroscopy} of the excitonic binding energy on magnetic field. Here $E_0$ is the ground state energy without the magnetic field, $B_c=2\hbar/(q\langle\rho_x^2\rangle)$ corresponds to the magnetic field when $2l_B^2=\langle\rho_x^2\rangle$, $\rho_{x}$ is the excitonic
Bohr radius, $l_B$ is the magnetic length, $\mu$ is the reduced effective mass, and $q$ the magnitude of the electronic charge. The fit to Eq.\ref{eqn: FIPS}, along with its
physical content, is also shown in Fig.\ref{fig: Fig1}. The second term in Eq.\ref{eqn: FIPS}(b) corresponds to the excitonic binding energy and is the extrapolation of the high-field slope [third term in Eq.\ref{eqn:
FIPS}(b)] to $B=0$.\cite{binding energy intecept} The crossover from  quadratic to linear slope is found at $5.9$T. The values of the exciton radius, the binding energy and the diamagnetic shift coefficient are given in Table 1.

\begin{figure}[!b]
\begin{center}
\includegraphics[width=8.5cm,
  keepaspectratio,
  angle=0,
  origin=lB]{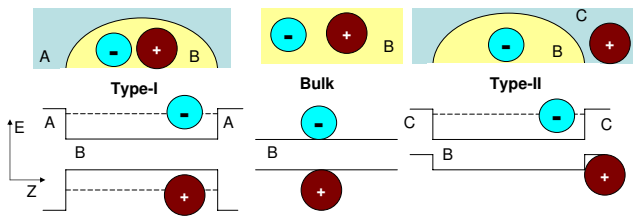}
\caption{\label{fig:band diagram}{(Color online) Schematic depiction of an exciton
in bulk material, and heterostructured QDs with type-I and type-II
band alignment. Material combination `B/A' has type-I alignment (e.g.
InAs/InP and possibly InP/InGaP) whereas `B/C' has
type-II band alignment (e.g. InP/GaAs). Type-I confinement results
in a decrease in the exciton radius and an increase in the exciton
binding energy. For QDs with type-II band alignment, the exciton
binding should be weaker and the radius larger compared to the
bulk InP values and the emission energy can be smaller than the
bandgaps of either of the two materials. }}
\end{center}
\end{figure}

Table 1 also compares the effect of the heterostructure boundary conditions on the excitonic parameters in InP. Note the systematic trend in the value of the diamagnetic shift coefficient. The effective electron-hole interaction in type-II dots is the weakest followed by bulk InP, and then finally InP/Ga$_x$In$_{1-x}$P QDs which show a much stronger binding [also see Fig.\ref{fig:band diagram}]. Thus, on physical grounds, the observations are most consistent with type-II band alignment.

\begin{table}[!b]
\caption{Comparison of exciton parameters in InP under different boundary conditions} 
\centering
\begin{tabular}{c c c c}
  \hline
  \hline
  $\,$ & Diamag. shift & Bohr radius & Binding energy\\               
  $\,$ & $\Gamma$ ($\mu eV T^{-2}$)& $\sqrt{\langle \rho^2\rangle}$ (nm) & $E_B$ (meV) \\       
  \hline
  InP/GaAs QD\footnote{This work}  & $42.4\pm 0.5$ & $15.0\pm 0.1$  &$1.5\pm 0.1$ \\

  Bulk InP & 40 [\onlinecite{godoy rsi}]& $\sim 12$ \footnote{Estimated from the diamagnetic shift in the first column using electron mass of InP, $\mu=0.08m_0$} & $4.8 \pm 0.2$ [\onlinecite{nam}]\\

  InP/GaInP QD & 2-5 [\onlinecite{Sugisaki_InP type1, provoost}]& $2.7-4.3$ {$^b$} & ---\footnote{Available magnetic field was not large enough\cite{provoost} to reach $B>B_c$ [cf. equation \ref{eqn: FIPS}(b)]}\\
\hline
\hline
\end{tabular}
\end{table}

\begin{figure}[!b]
\begin{center}
\includegraphics[width=9.3cm,
  keepaspectratio,
  angle=0,
  origin=lB]{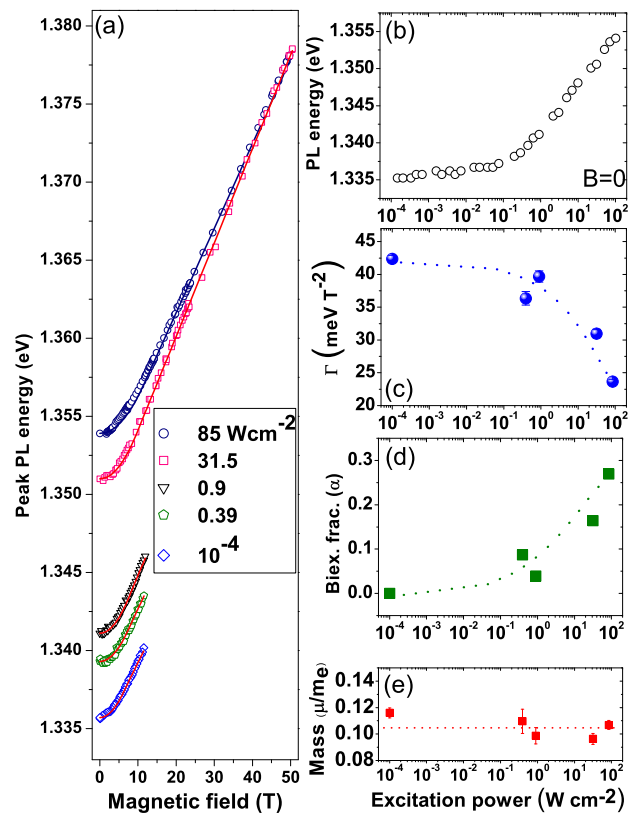} \caption{\label{fig:combined figure}{(Color online)(a) The magnetic field-dependent shift of the emission peak at different excitation powers--- 0.0001, 0.39, 0.9, 32, and 85 \wcm. Pulsed field measurements, which allow for only a short photon accumulation time, could only be performed at high powers. (b) Dependence of the PL energy on the excitation power measured at 4.2K (B=0). (c) The diamagnetic shift coefficient ($\Gamma$). (d) The biexciton fraction roughly inferred from (c) using equation 2. $\rho_{xx}/\rho_x$ was assumed to be 0.6, which is close to the ratio of hydrogen to helium radii. (e) The exciton reduced mass. The dotted lines in (c), (d) and (e) are just to guide the eye. }}
\end{center}
\end{figure}
Next, we will explore the multi-particle states in these QDs by excitation-power-dependent PL measurements. These measurements will also help us rule out DA recombination, as will be discussed in section III C. In what follows, we will assume that the QDs have type-II alignment.

\subsection{Biexcitons}
The PL peak position, measured without the magnetic field [Fig.\ref{fig:combined figure}(b)] is strongly excitation power-dependent beyond an incident laser intensity of about $10^{-1}$\wcm. This marks the point where multiparticle states start to play a role in PL. The observed blue-shift is due to the additional energy associated with the capacitive charging of the QDs. The integrated PL intensity (not shown here for brevity) also gradually changes its slope from linear to (slightly) superlinear beyond $10^{-1}$\wcm, indicating that a fraction of emission is from biexcitonic states.

The dependence of the PL peak position as a function of the magnetic field at different excitation powers, measured over almost six orders of magnitude, is shown in Fig.\ref{fig:combined figure}(a). Notice that the curves are all qualitatively similar and that they  all can be fitted to equations \ref{eqn: FIPS} [solid lines in Fig\ref{fig:combined figure}(a)]. This is because the B-dependent change in the energy of biexcitonic levels is just the sum of single particle energies, but with the important difference that the biexcitonic radius $\rho_{xx}$ and hence the diamagnetic shift coefficient are significantly different.

However, for an ensemble, the analysis is complicated by the facts that (i) the emission is from a mixture of excitonic and biexcitonic states with an unknown biexciton fraction $\alpha$, (ii) when a photon is emitted by a biexciton, what is measured in the magneto-PL experiment is the difference in the shifts of biexcitonic and excitonic levels, because the emission process involves $\textrm{biexciton}\rightarrow\textrm{exciton} + \textrm{photon}$. The measured change in the PL emission energy at low magnetic fields [diamagnetic shift regime, equation 1(a)] will then be
\begin{equation}\label{eqn:shift ensemble}
\Delta E_{\textrm{dia}}(B)={q^2\over
8\mu}\left[\alpha(2\langle\rho_{xx}^2\rangle-\langle\rho_x^2\rangle)+(1-\alpha)\langle\rho_{x}^2\rangle\right]B^2. \;\;\nonumber (2)
\end{equation}
The diamagnetic shift coefficient $\Gamma$ is thus expected to be strongly excitation-power-dependent. Indeed this is observed in Fig.\ref{fig:combined figure}(c). $\Gamma$ changes by factor of two due to the much smaller biexcitonic radius. Recall that in type-I QDs, confinement usually renders the diamagnetic shift independent of the charge in the dot.\cite{cade} In general, it depends on the relative extents of the spatial spread of the electron and the hole wave functions.

For a rough estimate of the biexciton fraction as a function of excitation power, we use the ratio of the Bohr radii for the helium and the hydrogen atom, $\rho_{\textrm He}/\rho_{\textrm H}\approx 0.6$.\cite{footnote1} Fig.\ref{fig:combined figure}(d) is plotted using this value of $\rho_{xx}/\rho_{x}$ and we find that the biexciton fraction at the highest excitation power is about 30\%. Also note that when $\langle\rho^2_{xx}\rangle/\langle\rho^2_x\rangle <0.5$, equation 2 predicts that the observed diamagnetic shift coefficient would not only reduce but also become negative at very high excitation powers.

The strong dependence of the diamagnetic shift coefficient on the excitation power also explains the anomalous results of Godoy, et al.\cite{godoy}, who found the excitonic diameter to be smaller in InP/GaAs compared to the bulk InP. They measured a diamagnetic shift coefficient of between 5--20$\mu eV T^{-2}$, the latter of which is equal to what we measure at the highest excitation power.

Fig.\ref{fig:combined figure}(e) shows that the exciton mass (high field slope of the curves in Fig.\ref{fig:combined figure}(a)) stays at approximately $0.1 m_e$, constant within 15\% over the whole range of excitation powers. Hence the changes in the diamagnetic shift coefficient can be understood as being largely the difference in the excitonic and biexcitonic radii  (equation 2). This provides consistency to the analysis. The value of the mass is between the free electron mass in InP ($0.08 m_e$) and the free heavy-hole in GaAs ($0.45m_e$). This is reasonable because the electron is largely immobilized by the quantum dot. Strain and non-parabolicity effects may further contribute to the enhancement of the electron mass.

\subsection{Donor-Acceptor-pair(DA)-recombination hypothesis}
The excitation power dependence of the diamagnetic shift coefficient [Fig.3(c)] also rules out recombination due to overlap between donor and acceptor wave functions. DA recombination has an energy\cite{klingshim}
\begin{equation}
\hbar \omega_{DA} = E_g-E^b_{D^0} -E^b_{A^0} + {e^2\over
4\pi\epsilon_0\epsilon r_{DA}} - m\hbar\omega_{LO}\;\;(3)\nonumber
\end{equation}
 $E_g$ is the energy gap, $E^b_{D^0}$ and $E^b_{A^0}$ are the binding energies of the donor and acceptor levels, the fourth term is the Coulomb repulsion energy of the ionized centres {\em after} recombination, and $m\hbar\omega_{LO}$ are the phonon-assisted transitions.\cite{klingshim} While the DA-pair emission shares some characteristics with emission from type-II QDs such as the diamagnetic shifts from DA-recombination would be of the same order as for excitons [Fig.1] and at higher excitation powers there would be a blue-shift\cite{klingshim} qualitatively similar to that seen in Fig.3(b), the diamagnetic shift coefficient itself should not have an excitation power dependence as observed in Fig.3(c). Secondly one should expect a strong quenching of the PL intensity in magnetic field for DA recombination---the wavefunctions shrink in the magnetic field and since
the electron and hole centres are not at the same point in space, their (exponentially small) overlap will be strongly reduced. This is also not observed.
\section{Conclusions}
We studied the PL from InP/GaAs QD heterostructures in very high magnetic fields ($B\leq 50T$) over
almost six orders of magnitude of excitation powers. Magneto-PL measurements at very low excitation powers established that the excitons have an average Bohr radius of $15$nm and a binding energy of $1.5$meV. These values indicate a much weaker binding for the excitons in comparison with bulk InP and provide strong evidence for type-II band alignment in these QDs.
We also studied the evolution of the electron-hole binding as the QD
ensemble makes a gradual transition from a regime where the
emission is from (hydrogen-like) two-particle excitonic states, to
a regime where the emission from (helium-like) four-particle
biexcitonic states also becomes significant. This was demonstrated by a strong variation of the diamagnetic shift with the excitation power.

\section{Acknowledgments}
The work at the KU Leuven is supported by the FWO and GOA programmes and by the Methusalem Funding of the Flemish Government.
MH acknowledges the support of the Research Councils UK.


\begin{references}
\bibitem{footnote1}A proper theoretical calculation would require solving the Schrodinger equation for the four-body problem in presence of suitable boundary conditions and magnetic field, which is beyond the scope of this work.
\end{references}
\end{document}